\documentclass[12pt]{iopart}

\newcommand{\msun}{~{\rm M}_\odot}
\def\lsim{\mathrel{\rlap{\lower4pt\hbox{\hskip1pt$\sim$}}
	\raise1pt\hbox{$<$}}} 
\def\gsim{\mathrel{\rlap{\lower4pt\hbox{\hskip1pt$\sim$}}
	\raise1pt\hbox{$>$}}} 
\def\be{\begin{eqnarray}}
\def\ee{\end{eqnarray}}
\usepackage{graphicx}
\begin{document}

\title[Quark matter and the astrophysics of neutron stars]
{Quark matter and the astrophysics of neutron stars }

\author{M Prakash}

\address{Department of Physics \& Astronomy, 
Ohio University, Athens, OH 45701, USA} 
\ead{prakash@harsha.phy.ohiou.edu}
\begin{abstract}
Some of the means through which the possible presence of nearly deconfined
quarks in neutron stars can be detected by astrophysical
observations of neutron stars from their birth to old age are
highlighted.
\end{abstract}


\section{Introduction}{\label{intro}}
Utilizing the asymptotic freedom of QCD, Collins and
Perry~\cite{Collins75} first noted that the dense cores of neutron
stars may consist of deconfined quarks instead of hadrons.  The
crucial question is whether observations of neutron stars from their
birth to death through neutrino, photon and gravity-wave emissions
can unequivocably reveal the presence of nearly
deconfined quarks instead of other possibilities such as 
only nucleons or other exotica such as strangeness-bearing hyperons or Bose
(pion and kaon) condensates.

\section{Neutrino signals during the birth of a neutron star}
The birth of a neutron star is heralded by the arrival of neutrinos on
earth as confirmed by IMB and Kamiokande neutrino detectors in the
case of supernova SN 1987A. Nearly all of the gravitational binding
energy (of order 300 bethes, where 1 bethe $\equiv 10^{51}$ erg)
released in the progenitor star's white dwarf-like core is carried off
by neutrinos and antineutrinos of all flavors in roughly equal
proportions.  The remarkable fact that the weakly interacting
neutrinos are trapped in matter prior to their release as a burst is
due to their short mean free paths in matter, $\lambda \approx (\sigma
n)^{-1} \approx 10$ cm, (here $\sigma \approx 10^{-40}~{\rm cm}^2$ is
the neutrino-matter cross section and $n \approx 2~{\rm to}~3~n_s$,
where $n_s \simeq 0.16~{\rm fm}^{-3}$ is the reference nuclear
equilibrium density), which is much less than the proto-neutron star
radius, which exceeds 20 km.  Should a core-collapse supernova occur
in their lifetimes, current neutrino detectors, such as SK, SNO,
LVD's, AMANDA, etc., offer a great opportunity for understanding a
proto-neutron star's birth and propagation of neutrinos in dense
matter insofar as they can detect tens of thousands of neutrinos in
contrast to the tens of neutrinos detected by IMB and Kamiokande.

\begin{figure}
\begin{center}
\includegraphics[height=275pt,angle=0]{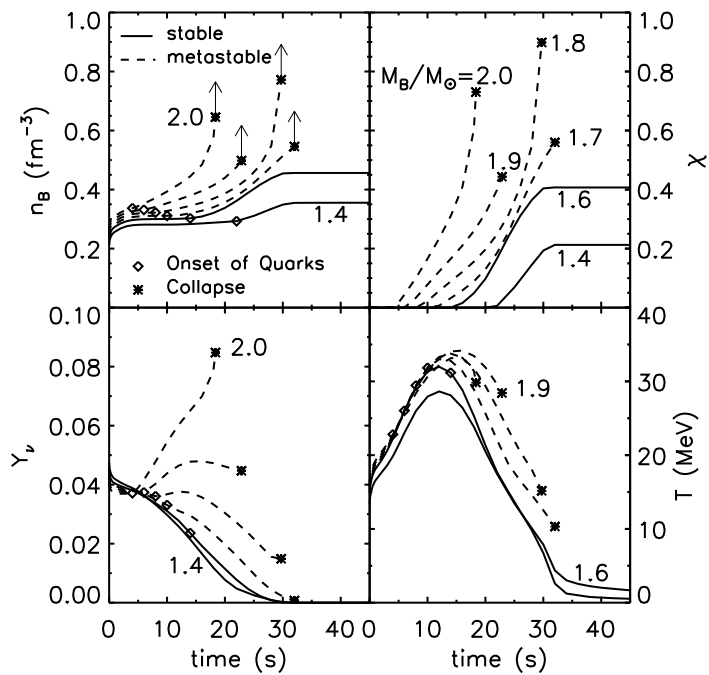}
\caption{Evolutions of the central baryon density $n_B$, $\nu$
  concentration $Y_\nu$, quark volume fraction $\chi$ and temperature
  $T$ for different baryon masses $M_B$. Solid lines show stable stars
  whereas dashed lines showing stars with larger masses are
  metastable. Diamonds indicate when quarks appear at the star's
  center, and asterisks denote when metastable stars become
  gravitationally unstable. Figure after Ref.~\cite{Pons01}}
\label{qevol}
\end{center}
\end{figure}

\begin{figure}
\begin{center}
\includegraphics[height=275pt,angle=0]{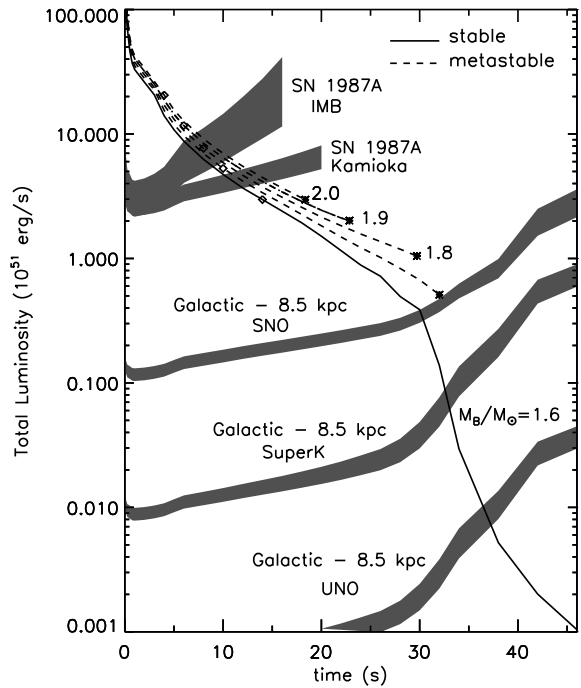}
\caption{The evolution of the total neutrino luminosity for stars of
  indicated baryon masses. Shaded bands illustrate the limiting
  luninosities corresponding to a count rate of 0.2 Hz in all
  detectors assuming 50 kpc for IMB and and Kamioka, 8.5 kpc for SNO,
  SuperK, and UNO. Shaded regions represent uncertanities in the
  average neutrino energy from the use of a diffusion scheme for
  neutrino transport in matter. Figure after Ref. \cite{Pons01}.}
\label{qpns_fig3}
\end{center}
\end{figure}

The appearence of quarks inside a neutron star leads to a
decrease in the maximum mass that matter can support, implying
metastability of the star. This would occur if the proto-neutron
star's mass, which must be less than the maximum mass of the hot,
lepton-rich matter is greater than the maximum mass of hot,
lepton-poor matter. For matter with nucleons only, such a
metastability is denied (see, e.g., \cite{ELP}).
Figure \ref{qevol} shows the evolution of some thermodynamic
quantities at the center of stars of various fixed baryonic masses.
With the equation of state used (see \cite{Pons01} for details), stars
with $M_B \lsim 1.1 \msun$ do not contain quarks and those with $M_B
\sim 1.7 \msun$ are metastable. The subsequent collapse to a black
hole could be observed as a cessation in the neutrino signals well
above the sensitivity limits of the current detectors (Figure
\ref{qpns_fig3}).

\section{Photon signals during the thermal evolution of a neutron star}

\begin{figure}
\begin{center}
\includegraphics[height=275pt,angle=90]{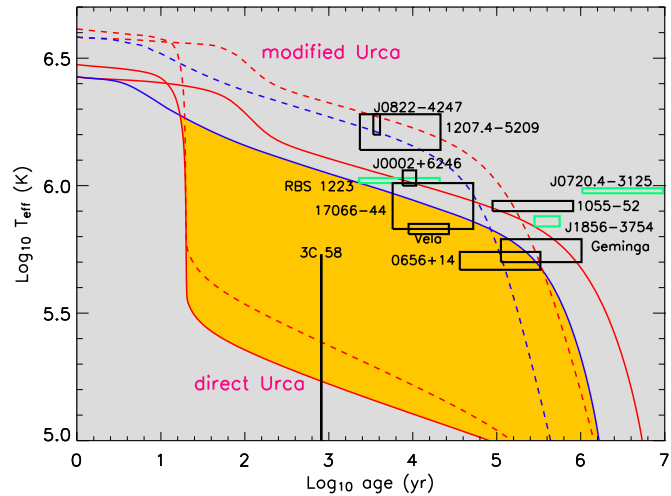}
\caption{Observational estimates of neutron star temperatures and ages
together with theoretical cooling simulations for $M=1.4$
M$_\odot$. Models and data are described in
\cite{page04}. Orange error boxes (see online) indicate sources from which both
X-ray and optical emissions have been observed. Simulations are for
models with Fe or H envelopes, with and without the effects of
superfluidity, and allowing or forbidding direct Urca
processes. Models forbidding direct Urca processes are relatively
independent of $M$ and superfluid properties. Trajectories for models
with enhanced cooling (direct Urca processes) and superfluidity lie within the
yellow region, the exact location depending upon $M$ as well as superfluid
and Urca properties. Figure adapted from Ref. \cite{LP04}.}
\label{cool_science}
\end{center}
\end{figure}

Multiwavelength photon observations of neutron stars, the bread and
butter affair of astronomy, has yielded estimates of the surface
tempeartures and ages of several neutron stars
(Fig. \ref{cool_science}). As neutron stars cool principally through
neutrino emission from their cores, the possibility exists that the
interior composition can be determined. The star continuosly emits
photons, dominantly in x-rays, with an effective temperature $T_{eff}$
that tracks the interior temperature but that is smaller by a factor
$\sim 100$.  The dominant neutrino cooling reactions are of a general
type, known as Urca processes \cite {LPPH}, in which thermally excited
particles undergo beta and inverse-beta decays. Each reaction produces
a neutrino or anti-neutrino, and thermal energy is thus continuously
lost.  Depending upon the proton-fraction of matter, which in turn
depends on the nature of strong interactions at high density, direct
Urca processes involving nucleons, hyperons or quarks lead to enhanced
cooling compared to modified Urca processes in which an additional
particle is required to conserve momentum. However, effects of
superfluidity abates cooling as sufficient thermal energy is required
to break paired fermions. In addition, the poorly known envelope
composition also plays a role in the inferred surface temperature
(Fig. \ref{cool_science}). The multitude of high density phases,
cooling mechanisms, effects of superfluidity, and unknown envelope composition
have thus far prevented definitive conclusions to be drawn (see, e.g.,
\cite{PPLS}).

\section{Mesured masses and their implications}
\begin{figure}
\begin{center}
\includegraphics[height=275pt,angle=90]{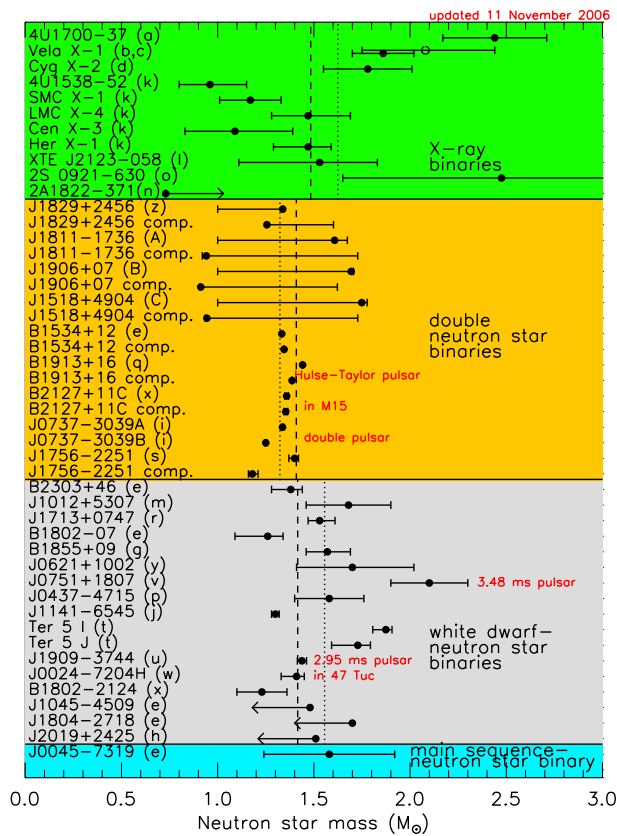}
\caption{Measured and estimated masses of neutron stars in binarie
  pulsars (gold, silver and blue regions online) and in x-ray
  accreting binaries (green). For each region, simple averages are
  shown as dotted lines; error weighted averages are shown as dashed
  lines. For labels and other details, consult Ref.~\cite{LP06}.}
\label{masses}
\end{center}
\end{figure}

Several recent observations of neutron stars have direct bearing on
the determination of the maximum mass.  The most accurately measured
masses are from timing observations of the radio binary pulsars. As
shown in Fig. \ref{masses}, which is compilation of the measured
neutron star masses as of November 2006, observations include pulsars
orbiting another neutron star, a white dwarf or a main-sequence star.

One significant development concerns mass determinations in binaries
with white dwarf companions, which show a broader range of neutron
star masses than binary neutron star pulsars. Perhaps a rather narrow
set of evolutionary circumstances conspire to form double neutron star
binaries, leading to a restricted range of neutron star masses
\cite{Bethe98}.  This restriction is likely relaxed for other neutron
star binaries.  A few of the white dwarf binaries may contain neutron
stars larger than the canonical 1.4 M$_\odot$ value, including the
intriguing case \cite{Nice05} of PSR J0751+1807 in which the estimated
mass with $1\sigma$ error bars is $2.1\pm0.2$ M$_\odot$. In addition,
to 95\% confidence, one of the two pulsars Ter 5 I and J has a
reported mass larger than 1.68 M$_\odot$ \cite{Ransom05}.

Whereas the observed simple mean mass of neutron stars with white
dwarf companions exceeds those with neutron star companions by 0.25
M$_\odot$, the weighted means of the two groups are virtually the
same.  The 2.1 M$_\odot$ neutron star, PSR J0751+1807, is about
$4\sigma$ from the canonical value of 1.4 M$_\odot$.
{\em It is furthermore the case that the $2\sigma$ errors of all but
two systems extend into the range below 1.45 M$_\odot$, so
caution should be exercised before concluding that firm evidence of
large neutron star masses exists.} Continued observations,
which will reduce the observational errors, are necessary to clarify
this situation.

Masses can also be estimated for another handful of binaries which
contain an accreting neutron star emitting x-rays.
Some of these systems are characterized by
relatively large masses, but the estimated errors are also large. The
system of Vela X-1 is noteworthy because its lower mass limit (1.6 to
1.7M$_\odot$) is at least mildly constrained by
geometry \cite{Quaintrell03}.

Raising the limit for the neutron star
maximum mass could eliminate entire families of EOS's, especially
those in which substantial softening begins around 2 to 3$n_s$. This
could be extremely significant, since exotica (hyperons, Bose
condensates, or quarks) generally reduce the maximum mass appreciably.

\section*{Ultimate energy density of observable cold baryonic matter}
\begin{figure}
\begin{center}
\includegraphics[height=275pt,angle=90]{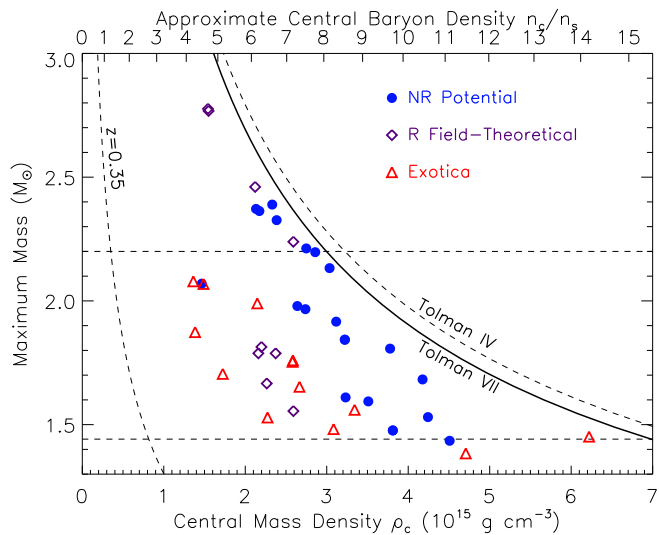}
\caption{
Model predictions are compared with results from the Tolman IV and
VII analytic solutions of general relativistic stucture equations. NR
refers to non-relativistic potential models, R are field- theoretical
models, and Exotica refers to NR or R models in which strong softening
occurs, due to hyperons, a Bose condensate, or quark
matter as well as self-bound strange quark matter. Constraints
from a possible redshift measurement of $z=0.35$ is also shown.  The
dashed lines for 1.44 and 2.2 M$_\odot$ serve to guide the eye. Figure
taken from Ref.~\cite{LP05}.}
\label{mmax}
\end{center}
\end{figure}

Measurements of neutron star masses can set an upper limit to the
maximum possible energy density in {\it any} compact object.  It has
been found \cite{LP05} that no causal EOS has a central density, for a
given mass, greater than that for the Tolman VII \cite{Tolman39}
analytic solution.  This solution corresponds to a quadratic
mass-energy density $\rho$ dependence on $r$,
$\rho=\rho_c[1-(r/R)^2]$, where the central density is $\rho_c$.  For
this solution,
\be\label{tolman}
\rho_{c,T~VII}=2.5\rho_{c,Inc}\simeq1.5\times10^{16}
\biggl({{\rm M}_\odot\over M}\biggr)^2 {\rm~g~cm}^{-3}\,.
\ee
A measured mass of 2.2 M$_\odot$ would imply
$\rho_{max}<3.1\times10^{15}$ g cm$^{-3}$, or about
$8n_s$. 

Figure \ref{mmax} displays maximum masses and accompanying central
densities for a wide wariety of neutron star EOS's, including models
containing significant softening due to ``exotica'', such as strange
quark matter.
The upper limit to the density could be lowered if the causal
constraint is not approached in practice.  For example, at high
densities in which quark asymptotic freedom is realized, the sound
speed is limited to $c/\sqrt{3}$.  Using this as a strict limit at all
densities, the Rhoades \& Ruffini \cite{Rhoades74} mass limit is
reduced by approximately $1/\sqrt{3}$ and the compactness limit
$GM/Rc^2=1/2.94$ is reduced by a factor $3^{-1/4}$ to $1/3.8$
\cite{Lattimer90}.  In this extreme case, the maximum density
would be reduced by a factor of $3^{-1/4}$ from that of Eq. (\ref{tolman}).
A 2.2 M$_\odot$ measured mass would imply a maximum density of about 
$4.2n_s$.

\section{Gravitational wave signals during mergers of binary stars}
Mergers of compact objects in binary systems, such as a pair of
neutron stars (NS-NS), a neutron star and a black hole (NS-BH), or two
black holes (BH-BH), are expected to be prominent sources of
gravitational radiation \cite{THORNE1}.  The gravitational-wave
signature of such systems is primarily determined by the chirp mass
$M_{chirp}=(M_1M_2)^{3/5}(M_1+M_2)^{-1/5}$, where $M_1$ and $M_2$ are
the masses of the coalescing objects. The radiation of gravitational
waves removes energy which causes the mutual orbits to decay. For
example, the binary pulsar PSR B1913+16 has a merger timescale of
about 250 million years, and the pulsar binary PSR J0737-3039 has a
merger timescale of about 85 million years \cite{Lyne04}, so there is
ample reason to expect that many such decaying compact binaries exist
in the Galaxy.  Besides emitting copious amounts of gravitational
radiation, binary mergers have been proposed as a source of the
r-process elements \cite{LSCH2} and the origin of the
shorter-duration gamma ray bursters \cite{EICHLER89}.

\begin{figure}
\begin{center}
\includegraphics[height=275pt,angle=90]{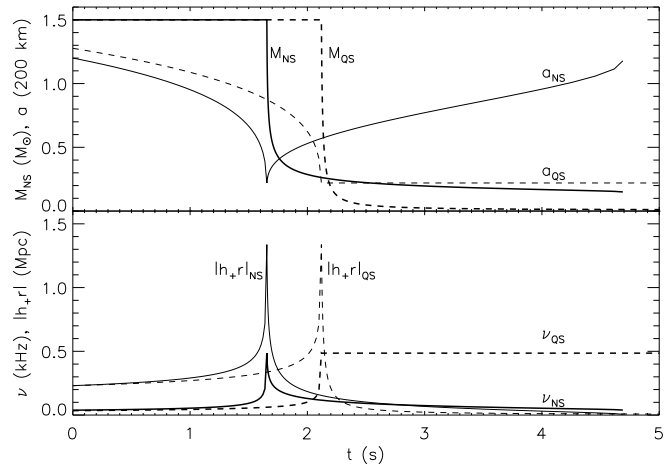}
\vspace*{-20pt}
\caption{Physical and observational
  variables in mergers between low-mass black holes and neutron stars
  or self-bound quark stars.  The total system mass is 6 M$_\odot$ and
  the initial mass ratio is $q=1/3$ in both cases.  The initial radii
  of the neutron star and quark star were assumed to be equal.  The
  time scales have arbitrary zero points.  Upper panel displays
  semi-major axis $a$ (thick lines) and component mass $M_{NS},
  M_{QS}$ (thin lines) evolution.  Lower panel displays orbital
  frequency $\nu$ (thick lines) and strain amplitude $|h_+r|$
  evolution.  Solid curves refer to the neutron star
  simulation and dashed curves to the quark star simulations. Figure
  taken from Ref. \cite{LP06}.}
\label{gwevol} 
\end{center}
\end{figure}

Observations of gravity waves from merger events can simultaneously
measure masses and radii of neutron stars, and could set firm limits
on the neutron star maximum mass \cite{PL03,RPL05}. 
Binary mergers for the two cases of a black hole and a normal neutron
star and a black hole and a self-bound strange quark matter star
(Fig. \ref{gwevol})
illustrate the unique opportunity afforded by gravitational wave
detectors due to begin operation over the next decade, including LIGO,
VIRGO, GEO600, and TAMA.   

A careful analysis of the
gravitational waveform during inspiral yields values for not only the
chirp mass $M_{chirp}$, but for also the
reduced mass $M_{BH}M_{NS}/M$, so that both $M_{BH}$ and $M_{NS}$ can
be found \cite{CUTLER94}.  
The onset of mass transfer can be determined by the peak in $\omega$,
and the value of $\omega$ there gives $a$.  
A general relativistic analysis of mass transfer conditions then
allows the determination of the star's radius \cite{RPL05}.  Thus a
point on the mass-radius diagram can be estimated \cite{FABER02}.
The combination $h_+\omega^{-1/3}$ depends only on a function of $q$,
so the ratio of that combination and knowledge of $q_i$ should allow
determination of $q_f$.  From the Roche condition and knowledge of
$a_f$ from $\omega_f$,  another mass-radius combination can be found.

The sharp contrast between the evolutions during stable mass transfer
of a normal neutron star and a strange quark star should make these
cases distinguishable.  For strange quark matter stars, the
differences in the height of the frequency peak and the plateau in the
frequency values at later times are related to the differences in
radii of the stars at these two epochs.  It could be an indirect
indicator of the maximum mass of the star: the closer is the star's
mass before mass transfer to the maximum mass, the greater is the
difference between these frequency values, because the radius change
will be larger.  Together with radius information, the value of the
maximum mass remains the most important unknown that could reveal the
true equation of state at high densities.

\ack
This work was supported in part by the U.S. Department of Energy under
the grant DOE/DE-FG02-93ER40756.

\end{document}